\documentclass[12pt,a4paper,english,showpacs,pra,superscriptaddress]{revtex4}
\usepackage[T1]{fontenc}
\usepackage[latin9]{inputenc}
\usepackage{babel}
\usepackage{amsmath}
\usepackage{amssymb}
\usepackage{esint}
\usepackage[unicode=true,pdfusetitle,
 bookmarks=true,bookmarksnumbered=false,bookmarksopen=false,
 breaklinks=false,pdfborder={0 0 1},backref=false,colorlinks=false]
 {hyperref}
\usepackage{breakurl}

\makeatletter

\@ifundefined{textcolor}{}
{%
 \definecolor{BLACK}{gray}{0}
 \definecolor{WHITE}{gray}{1}
 \definecolor{RED}{rgb}{1,0,0}
 \definecolor{GREEN}{rgb}{0,1,0}
 \definecolor{BLUE}{rgb}{0,0,1}
 \definecolor{CYAN}{cmyk}{1,0,0,0}
 \definecolor{MAGENTA}{cmyk}{0,1,0,0}
 \definecolor{YELLOW}{cmyk}{0,0,1,0}
}

\usepackage{babel}

\makeatother

\begin{document}

\title{Engineering the unitary charge-conjugation operator of quantum field
theory for particle-antiparticle using trapped ions and light fields
in cavity QED}

\author{N.G. de Almeida}

\affiliation{Instituto de Física, Universidade Federal de Goiás, 74.001-970, Goiânia
(GO), Brazil}

\affiliation{Departamento de Fisica Fundamental and IUFFyM, Universidad de Salamanca,
Spain}
\begin{abstract}
We present a method to engineer the unitary charge conjugation operator,
as given by quantum field theory, in the highly controlled context
of quantum optics, thus allowing one to simulate the creation of charged
particles with well-defined momenta simultaneously with their respective
antiparticles. Our method relies on trapped ions driven by a laser
field and interacting with a single mode of a light field in a high
Q cavity.
\end{abstract}

\pacs{42.50.Ct, 32.80.-t, 42.50.Dv,42.50.Pq,}

\maketitle

\section{Introduction}

\paragraph*{\textup{The charge-conjugation, together with parity and time reversal
operators, gives rise to one of the most profound discrete symmetry
in nature. Charge-conjugation operator replaces the charged field
by their charge-conjugate fields, or, equivalently, transform a particle
into its corresponding antiparticle. Parity operator, on the other
hand, produces a space reflexion on the particle and antiparticle
states of the field, while time reversal, in turn, act on the Hilbert
space in the same form as the parity operator. These charge, parity,
and time reversal (CPT) symmetries each is known to be valid for free
(noninteracting) fields, and culminate in the CPT theorem \cite{Greiner book,Peskin book},
which states that a local and Lorentz-covariant quantum field theory
is invariant under the combined CPT operation \cite{Greenberg02}.
Take as an example the Klein-Gordon field, describing a field with
electrical charged particles of spin zero. In order to explain the
emergence of particles charged electrically, we must resort to complex
fields, such that, for example, the Hermitian charge operator as well
as the charge conjugation and parity operators can be build \cite{Ryder book,Greiner book}.
In this paper we show how to engineer an operator similar to the unitary
charge-conjugation operator of two modes of both the complex Klein-Gordon
and Dirac fields in the context of trapped ions \cite{Meekhof96},
thus opening the possibility to simulate creation and annihilation
of particle and antiparticles in a very controllable scenario.}}

\section{The Theory}

Although well known in quantum field theory, let us begin, for clarity,
with a brief review of the main results used here. The charge conjugation
is a symmetry of the theory, as for example, given a (density) Lagrangian
$\mathcal{L\textrm{(x)}}$ and a unitary charge conjugation operator
$\mathcal{\mathit{C}}$, then $\mathcal{\mathit{C^{-1}}L\textrm{(x)}C=L\textrm{(x)}}$. 

To be specific, consider first \cite{Greiner book,Ryder book} the
complex Klein-Gordon field for spin zero particles

\begin{equation}
\mathcal{L}\textrm{(x)}=(\partial_{\mu}\phi)(\partial^{\mu}\phi^{*})-m^{2}\phi\phi^{*}
\end{equation}
where $\phi$ and $\phi^{*}$ are considered independent complex fields,
and $m$ is the mass of the particle associated to the field excitations.
The Euler-Lagrange equations lead to the Klein-Gordon equation

\begin{equation}
(\partial_{\mu}\partial^{\mu}+m^{2})\phi=0.
\end{equation}
 In field quantization, the complex fields $\phi$ and $\phi^{*}$
are regarded as operators, such that ($x\equiv(x,y,z),k\equiv(k_{x},k_{y},k_{z})$)

\begin{equation}
\phi(x,t)=\sum_{k}[a_{k}u_{k}(x,t)+b_{k}^{\dagger}u_{k}^{*}(x,t)],
\end{equation}

\begin{equation}
\phi^{\dagger}(x,t)=\sum_{k}[a_{k}^{\dagger}u_{k}^{*}(x,t)+b_{k}u_{k}(x,t)],
\end{equation}
are the field operators replacing the complex scalar fields. The functions
$u_{k}(x,t)$, and $u_{k}^{*}(x,t)$ are box normalized imposing periodic
boundary conditions at a surface of a cube of volume $V$ such that
$(u_{j},u_{l})=\delta_{jl}$, $j,l=(k_{x},k_{y},k_{z})$, and imposing
commutation relations to the the bosonic operators $a_{k},$ $a_{k}^{\dagger}$,
and $b_{k}$, $b_{k}^{\dagger}$. Both $\mathcal{L}\textrm{(x)}$
in Eq.(1) and Eq.(2) are invariant under the gauge transformation
of first kind $\phi\rightarrow\phi exp(i\Lambda)$, and so by the
Noether's theorem there is a conserved current $j^{\mu}$and a conserved
charge $Q=\int j^{0}d^{3}x$ , whose density is given by $\mathit{\mathcal{Q}}=-i:\phi^{\dagger}\frac{\partial\phi}{\partial t}-\phi\frac{\partial\phi^{\dagger}}{\partial t}:$,
and $::$ denotes normal ordering. Using Eqs.(3)-(4) and the definition
$(\psi,\chi)\equiv i\int d^{3}x\psi^{*}\overleftrightarrow{\partial_{0}}\chi=i\int d^{3}x(\partial_{0}\psi^{*}\chi-\psi^{*}\partial_{0}\chi)$
, the charge operator can be written as

\begin{equation}
Q=\sum_{k}(a_{k}^{\dagger}a_{k}-b_{k}^{\dagger}b_{k}).
\end{equation}
Note that, given an eigenstate $\left\vert q\right\rangle $ of $Q$
with eigenvalue $q$, then $\phi^{\dagger}\left\vert q\right\rangle $
($\phi\left\vert q\right\rangle $) is also an eigenstate of $Q$
with eigenvalue $q+1$ ($q-1$), and, since $Q$ commutes with the
Hamiltonian, it is a conserved quantity, as should.

The charge-conjugation operator $C$ is defined to transform a particle
into its antiparticle, i.e.,

\begin{equation}
\mathit{C^{-1}}\phi(x,t)C=p\phi(x,t);C^{-1}\phi^{\dagger}(x,t)C=p^{*}\phi(x,t)
\end{equation}
which is equivalent to $\mathit{C^{-1}}a_{k}C=pb_{k}$, $C^{-1}a_{k}^{\dagger}C=p^{*}b^{\dagger}$,
and $C^{-1}b_{k}C=p^{*}a_{k}$, $C^{-1}b_{k}^{\dagger}C=pa$, with
$\left|p\right|=\pm1$. The charge-conjugation operator $C$ that
we want to simulate has the following unitary form\cite{Greiner book}

\begin{equation}
C=\exp\left[-\frac{i\pi}{2}\sum_{k}\left[a_{k}^{\dagger}b_{k}+b_{k}^{\dagger}a_{k}-p\left(a_{k}^{\dagger}a_{k}+b_{k}^{\dagger}b_{k}\right)\right]\right].
\end{equation}
Note that $Q$ anticomutes with $C$: $CQ=-QC$, and therefore in
general they do not possess the same eigenstate.

Consider now the Dirac field for particles of mass $m$ and spin $1/2$,
whose Lagrange density is

\begin{equation}
\mathcal{L}\textrm{(x)}=\frac{i}{2}\left[\overline{\psi}\gamma^{\mu}\left(\partial_{\mu}\psi\right)-\left(\partial_{\mu}\overline{\psi}\right)\gamma^{\mu}\psi\right]-m\overline{\psi}\psi\equiv\frac{i}{2}\overline{\psi}\gamma^{\mu}\overleftrightarrow{\partial}\psi-m\overline{\psi}\psi
\end{equation}
where $\gamma^{\mu}$, $\mu=0,1,2,3$ are the Dirac matrices and $\overline{\psi}=\psi^{\dagger}\gamma^{0}$.
The Euler-Lagrange equations now lead to the Dirac equation 

\begin{equation}
\left(i\gamma^{\mu}\partial_{\mu}-m\right)\psi=0,
\end{equation}
whose general solutions can be expanded in plane waves ($k\equiv(k_{x},k_{y},k_{z}$)
\begin{equation}
\psi(x,t)=\sum_{k,s}[c_{k,s}u_{k,s}\exp-i(kx-\omega_{k}t)+d_{k,s}^{\dagger}v_{k,s}\exp i(kx-\omega_{k}t)],
\end{equation}

\begin{equation}
\psi^{\dagger}(x,t)=\sum_{k,s}[c_{k,s}^{\dagger}u_{k,s}^{\dagger}\exp i(kx-\omega_{k}t)+d_{k,s}v_{k,s}^{\dagger}\exp-i(kx-\omega_{k}t)],
\end{equation}
where $s=1,2$ denotes the covariantly generalized spin vector \cite{Greiner book,Ryder book}
for the orthogonal energy positive $u_{k,s}(x,t)$ and energy negative
$v_{k,s}(x,t)$ spinors, and the fermionic operators $c_{k,s},$$c_{k,s}^{\dagger}$
and $d_{k,s},$$d_{k,s}^{\dagger}$ obeys anticomutation relations,
being interpreted as creator and annihilator of particles and antiparticles.
Using Eq.(10)-(11) and the orthogonality relations for the spinors:
$(u_{k,s}^{\dagger},u_{k,s^{'}})=(v_{k,s}^{\dagger},v_{k,s^{'}})=\delta_{ss^{'}}$
and $(u_{-k,s}^{\dagger},v_{k,s^{'}})=(v_{-k,s}^{\dagger},u_{k,s^{'}})=0$,
one finds for the charge operator $Q=\int j^{0}d^{3}x=e\int d^{3}x\psi^{\dagger}(x,t)\psi(x,t)$ 

\begin{equation}
Q=e\sum_{k,s}(c_{k,s}^{\dagger}c_{k,s}-d_{k,s}^{\dagger}d_{k,s}),
\end{equation}
where the elementary charge $e$ was explicitly inserted in the current
density vector $j_{\mu}=e\overline{\psi}(x,t)\gamma_{\mu}\psi(x,t)$.
Now, similarly to the spinless case, requiring that the charge-conjugation
operator $C$ , besides being unitary, transforms a particle into
its antiparticle as

\begin{equation}
\mathit{C^{-1}}c_{k}C=d_{k},\mathit{C^{-1}}c_{k}^{\dagger}C=d_{k}^{\dagger},
\end{equation}

\begin{equation}
\mathit{C^{-1}}d_{k}C=c_{k},\mathit{C^{-1}}d_{k}^{\dagger}C=c_{k}^{\dagger},
\end{equation}
the $C$ operator, which we want to engineer, reads\cite{Greiner book}

\begin{equation}
C=\exp-\frac{i\pi}{2}\sum_{k,s}\left[d_{k,s}^{\dagger}c_{k,s}+c_{k,s}^{\dagger}d_{k,s}-c_{k,s}^{\dagger}c_{k,s}-d_{k,s}^{\dagger}d_{k,s}\right].
\end{equation}

\section{Engineering the Unitary Charge-Conjugation Operator for Klein-Gordon
and Dirac Fields}

For our purpose, we consider just one pair of particle antiparticle,
such that the sum in Eqs.(7) disappears and we are lead essentially
with Hamiltonian of the type $H_{CC}=\frac{\pi}{2}\left[a^{\dagger}b+b^{\dagger}a-p\left(a^{\dagger}a+b^{\dagger}b\right)\right]$,
which, in the interaction picture, reads $H_{INT}=\frac{\pi}{2}\left(a^{\dagger}b+b^{\dagger}a\right)$.
Note that the Hamiltonian $H_{CC}$ producing the particle anti-particle
charge-conjugation operation can be combined in a single Hamiltonian
given by $H_{I}=ga^{\dagger}b+g*b^{\dagger}a$, provided that we choose
$g=\left|g\right|\exp(i\pi/2)$. Now, consider a single two-level
ion of mass $m$ whose frequency of transition between the excited
state $\left|e\right\rangle $ and the ground state $\left\lfloor g\right\rangle $
is $\omega_{0}.$ This ion is trapped by a harmonic potential of frequency
$\nu$ along the axis $x$ and driven by a laser field of frequency
$\omega_{l}$. The laser field promotes transitions between the excited
and ground states of the ion through the dipole constant $\Omega=\left|\Omega\right|\exp(i\phi_{l})$.
Finally, the ion is put inside a cavity containing a single mode of
a standing wave field of frequency $\omega_{f}$, such that the Hamiltonian
for this system reads $H=H_{0}+H_{1}$, where ($\hbar=1$)

\begin{equation}
H_{0}=\omega_{f}a^{\dagger}a+\hbar\nu b^{\dagger}b+\frac{\omega_{0}}{2}\sigma_{z}
\end{equation}

\begin{equation}
H_{1}=\sigma_{eg}\Omega\exp[i(k_{l}x-\omega_{l}t)]+\lambda a\sigma_{eg}cos(k_{f}x)+h.c.
\end{equation}
Here, $h.c$. is for hermitian conjugate, $a^{\dagger}$($a$) is
the creation (annihilation) operator of photons for the cavity mode
field and and $b^{\dagger}$ ($b$) is the corresponding creation
(annihilation) operator of phonons for the ion vibrational center-of-mass.
The ion center-of-mass position operator is $x=\frac{1}{\sqrt{2m\nu}}(b^{\dagger}+b)$,
while $\sigma_{z}=\left|e\right\rangle \left\langle e\right|-\left|g\right\rangle \left\langle g\right|$
and $\sigma_{ij}=\left|i\right\rangle \left\langle j\right|$, $i,j=\left\{ e,g\right\} $are
the Pauli operators.

Next, we consider the so-called Lamb-Dick regime, $\eta_{\alpha}=\bar{n}k_{\alpha}/\sqrt{1/2m\nu}\ll1$,
$\alpha=l,f$ and $\bar{n_{\alpha}}$ is the average photon/phonon
number. Thus, in the interaction picture and after discarding the
counter rotating terms in the so-called rotating wave approximation
(RWA), assuming that $\omega_{0}-\omega_{l}\approxeq\nu$ and $\omega_{0}\approxeq\omega_{f}$,
the Hamiltonian above reads

\begin{equation}
H_{RWA}=i\eta_{l}\Omega b\sigma_{eg}\exp[i(\omega_{0}-\omega_{l}-\nu)t]+\lambda a\sigma_{eg}\exp[i(\omega_{0}-\omega_{f})]+h.c.
\end{equation}

An effective Hamiltonian can be obtained from the above one by requiring
$\left|\eta_{l}\Omega\right|\sqrt{\left\langle b^{\dagger}b\right\rangle },\left|\lambda\right|\sqrt{\left\langle a^{\dagger}a\right\rangle }\ll\left|\omega_{0}-\omega_{l}-\nu\right|,\left|\omega_{0}-\omega_{f}\right|,$
and neglecting the highly oscillating terms stemming from\cite{James00}

\begin{equation}
H_{eff}=-iH(t)\int^{t}H(t')dt',
\end{equation}
where, in this notation, the lower limit is to be ignored. From Eq.(18)
the following effective Hamiltonian is obtained, provided the atom
internal state be prepared in the eigenstate of $\sigma_{gg}$:

\begin{equation}
H_{eff}=\omega_{a}a^{\dagger}a+\omega_{b}b^{\dagger}b+ga^{\dagger}b+g^{*}b^{\dagger}a,
\end{equation}
where $\omega_{a}=\left|\lambda_{a}\right|^{2}/(\omega_{0}-\nu)$;
$\omega_{b}=\eta^{2}\left|\Omega\right|^{2}/(\omega_{l}-\omega_{0})$;
$g=i\Omega\eta_{l}\lambda_{a}^{*}/(\omega_{0}-\nu)$ and an irrelevant
constant was disregarded. We can simplify further by choosing $\omega_{a}=\omega_{b}=\omega$
such that the effective Hamiltonian, after the unitary operation $U=\exp-i\omega t\left(a^{\dagger}a+b^{\dagger}b\right)$,
reads

\begin{equation}
H_{eff}=ga^{\dagger}b+g^{*}b^{\dagger}a.
\end{equation}
The desired charge-conjugation operation is obtained applying a laser
pulse of duration $\tau$ satisfying $g\tau=\pi/2\exp(i3\pi/2)$.
It is to be noted that the particle and antiparticle behavior is encoded
in the cavity-mode field, whose quanta creation is denoted by $a^{\dagger}$,
and in the vibrational motion of the ion center-of-mass, whose quanta
creation in turn is $b^{\dagger}$. 

The charge-conjugation operator engineering involving just two vibrational
center-of-mass motions along the axis $x$ and $y$ of a single ion
of mass $m$ can be attained in the following way. Consider the Hamiltonian
$H=H_{0}+H_{I}$ for a two-level ion constrained in a two-dimensional
harmonic trap driven by two traveling wave fields, characterized by
the two frequencies $\nu_{x}$ and $\nu_{y}$ propagating in $x$
and $y$ directions, respectively \cite{MoyaCessa12}, with

\begin{equation}
H_{0}=\nu_{x}a^{\dagger}a+\nu_{y}b^{\dagger}b+\frac{\omega_{0}}{2}\sigma_{z}
\end{equation}

\begin{equation}
H_{1}=\sigma_{eg}\Omega_{x}\exp[-i\left(k_{x}x-\omega_{x}t\right)]+\sigma_{eg}\Omega_{y}\exp[-i\left(k_{y}y-\omega_{y}t\right)]+h.c.,
\end{equation}
where $x=\sqrt{1/m\nu_{x}}(a+a^{\dagger})$ and $y=\sqrt{1/m\nu_{y}}(b+b^{\dagger})$
are the center-of-mass position operators of the ion in the $x-y$
plane, $\omega_{x}$ and $\omega_{y}$ are the frequencies of the
traveling fields of wave vectors $k_{x}$, $k_{y}$. As before, the
frequency of transition between the excited state $\left|e\right\rangle $
and the ground state $\left\lfloor g\right\rangle $ is $\omega_{0}$
and $\sigma_{ij}=\left|i\right\rangle \left\langle j\right|$, $i,j=g,e$.
In the interaction picture and assuming that $\eta_{x}=k_{x}\sqrt{1/m\nu_{x}},\eta_{y}=k_{y}\sqrt{1/m\nu_{y}}\ll1$,
the Hamiltonian above reads
\begin{align}
H_{INT} & =h.c.+\Omega_{x}\sigma_{eg}\exp\left[-i(\omega_{x}-\omega_{0})t\right]-i\eta_{x}a\sigma_{eg}\exp\left[-i(\omega_{x}-\omega_{0}+\nu_{x})t\right]-i\eta_{x}a^{\dagger}\sigma_{eg}\exp\left[-i(\omega_{x}-\omega_{0}-\nu_{x})t\right]\nonumber \\
 & +\Omega_{y}\sigma_{eg}\exp\left[-i(\omega_{y}-\omega_{0})t\right]-i\eta_{y}b\sigma_{eg}\exp\left[-i(\omega_{y}-\omega_{0}+\nu_{y})t\right]-i\eta_{y}b^{\dagger}\sigma_{eg}\exp\left[-i(\omega_{y}-\omega_{0}-\nu_{y})t\right]\label{2}
\end{align}

Now, by adjusting $\omega_{\alpha}\neq\omega_{0}$, $\alpha=x,y$,
such that $\left|\omega_{\alpha}-\omega_{0}\right|t\gg1$, $\omega_{x}-\omega_{0}\cong-\nu_{x}$
and $\omega_{y}-\omega_{0}\cong-\nu_{y}$, we can disregard the the
highly oscillating terms (RWA) such that in the weak coupling regime
where $\left|\Omega_{\alpha}\right|\sqrt{\overline{n}_{\alpha}}\ll\delta_{\alpha}$,
we are left with

\begin{equation}
H_{RWA}=-i\eta_{x}a\sigma_{eg}\exp\left(-i\delta_{x}t\right)+i\eta_{x}a^{\dagger}\sigma_{ge}\exp\left(i\delta_{x}t\right)-i\eta_{y}b\sigma_{eg}\exp\left(-i\delta_{y}t\right)+i\eta_{y}b^{\dagger}\sigma_{ge}\exp\left(i\delta_{y}t\right),
\end{equation}
and $\delta_{\alpha}=\omega_{\alpha}-\omega_{0}+\nu_{\alpha}\cong0$,
$\alpha=x,y$ . A particularly simple effective Hamiltonian is found
if we let $\delta_{x}=\delta_{y}$. In this case, using Eq.(12) we
found that the dynamics of the internal state decouples from the external
ones, such that $H_{total}=\sigma_{z}\oplus H_{eff}$, with

\begin{equation}
H_{eff}=\frac{\eta_{x}^{2}}{\delta}a^{\dagger}a+\frac{\eta_{y}^{2}}{\delta}b^{\dagger}b+\frac{\eta_{x}\eta_{y}}{\delta}\left(a^{\dagger}b+ab^{\dagger}\right).
\end{equation}
If we further choose $\nu_{x}=\nu_{y}$ or, equivalently, $\omega_{x}=\omega_{y}$,
then the Hamiltonian corresponding the charge-conjugation operator
Eq.(7) can be tailored adjusting $\frac{\eta^{2}}{\delta}\tau=\pm\frac{\pi}{2}$
in Eq.(26). 

Consider now the charge conjugation for fermions, Eq.(15). As previously
done for bosons, we will interested in just a single mode and a well
defined spin vector, such that we can write Eq.(15) as $C=\exp\left[-\frac{i\pi}{2}\left[\left(d^{\dagger}c+c^{\dagger}d\right)-\left(c^{\dagger}c+d^{\dagger}d\right)\right]\right]$,
with the particle and antiparticle operator obeying the anticomutation
relation. To engineer this Hamiltonian, let us consider now two two-level
ions $1$ and $2$, having the internal states described by pseudo-spin
operators $\sigma_{1}^{+}$,$\sigma_{1}^{-}$, $\sigma_{2}^{+}$,
$\sigma_{2}^{-}$ possessing the same algebra as those of $c^{\dagger}$,$c$,
$d^{\dagger}$ and $d$, respectively, while the one-dimension harmonic
motional states of each atom are described by the creation and annihilation
bosonic operators $b_{1}^{\dagger}$, $b_{1}$, $b_{2}^{\dagger}$,
and $b_{2}$. These two ions are put into the same cavity containing
a single mode of frequency $\omega_{a}$ of a electromagnetic standing
wave, such that the Hamiltonian for this system can be written as
$H=H_{0}+H_{1}$ , with ($\hbar=1$)

\begin{equation}
H_{0}=\omega_{a}a^{\dagger}a+\nu_{1}b_{1}^{\dagger}b_{1}+\nu_{2}b_{2}^{\dagger}b_{2}+\frac{\omega_{01}}{2}\sigma_{1}^{z}+\frac{\omega_{02}}{2}\sigma_{2}^{z}
\end{equation}

\begin{equation}
H_{1}=\lambda_{1}a\sigma_{1}^{+}\cos\left(\eta_{1}x_{1}\right)+\lambda_{2}a\sigma_{2}^{+}\cos\left(\eta_{2}x_{2}\right)+h.c.,
\end{equation}
where, in $H_{0}$, $a^{\dagger}$ and $a$ are the creation and annihilation
operator in Fock space for the cavity mode field, $\nu_{1}$ ($\nu_{2}$)
is the frequency of the ion trap $1$ ($2$) $\omega_{01}$ ($\omega_{02}$)
is the frequency of transition from the ground to the excited state
of ion $1$ ($2$), and, in $H_{1}$, $\eta_{\alpha}=k_{a}\sqrt{1/m\nu_{\alpha}},$
$\alpha=1,2$, is the Lamb-Dick parameter, and$\lambda_{1}$ ($\lambda_{2}$)
describes the strength of the coupling between the standing wave and
the ion placed at position $x_{1}$ ($x_{2}$) . 

Assuming that $\eta_{\alpha}\ll1$ and moving to the interaction picture,
the Hamiltonian above reads

\begin{equation}
H_{1}=\lambda_{1}a\sigma_{1}^{+}\exp\left[i\left(\omega_{o1}-\omega_{a}\right)t\right]+\lambda_{2}a\sigma_{2}^{+}\exp\left[i\left(\omega_{o2}-\omega_{a}\right)t\right]+h.c.
\end{equation}
For two identical ions, $\omega_{o1}=\omega_{o2}=\omega_{o}$; $\lambda_{1}=\lambda_{2}=\lambda$,
and under the assumption of weak coupling, $\left|\lambda_{\alpha}\right|\sqrt{\overline{n}}\ll\delta_{\alpha}=\left(\omega_{o\alpha}-\omega_{a}\right)$,
using Eq.(19) the following effective Hamiltonian is obtained: 
\begin{equation}
H_{eff}=\frac{\lambda^{2}}{\delta}\sigma_{1}^{+}\sigma_{1}^{-}+\frac{\lambda^{2}}{\delta}\sigma_{2}^{+}\sigma_{2}^{-}+\frac{\lambda^{2}}{\delta}a^{\dagger}a\left(\sigma_{1}^{z}+\sigma_{2}^{z}\right)+\frac{\lambda^{2}}{\delta}\left(\sigma_{1}^{+}\sigma_{2}^{-}+\sigma_{1}^{-}\sigma_{2}^{+}\right).
\end{equation}
If the system is tailored to fit $\frac{\lambda^{2}}{\delta}\tau=\pm\pi/2,$
and the cavity starts from the vacuum state, then the above Hamiltonian
gives rise to the desired charge-conjugation evolution operator 
\begin{equation}
C=\exp\left\{ -i\frac{\pi}{2}\left[\pm\left(\sigma_{1}^{+}\sigma_{1}^{-}+\sigma_{2}^{+}\sigma_{2}^{-}\right)\pm\left(\sigma_{1}^{+}\sigma_{2}^{-}+\sigma_{1}^{-}\sigma_{2}^{+}\right)\right]\right\} .
\end{equation}
which is similar to the charge-conjugation operator $C=\exp\left[-\frac{i\pi}{2}\left[\left(d^{\dagger}c+c^{\dagger}d\right)-\left(c^{\dagger}c+d^{\dagger}d\right)\right]\right]$. 

Before to finish this Section, let us briefly stress the feasibility
of our proposal. In order to engineer the Hamiltonians allowing to
simulate the particle and antiparticle charge-conjugation, we impose
the Lamb-Dick approximation, which consists in assuming the ion confined
within a region much smaller than the laser wavelength, such that
the we can safely choose the Lamb-Dick parameter as $\eta\sim0.1$
\cite{Meekhof96}. For the case of bosons, as for instance Eq.(21),
a one-dimensional trap is required, and we imposed the condition for
the laser frequencies $\omega_{a}=\omega_{b}=\omega$, which, in turns,
requires that $\left|\lambda_{a}\right|^{2}/(\omega_{0}-\nu)=\eta^{2}\left|\Omega\right|^{2}/(\omega_{l}-\omega_{0}).$
As for example, for a fixed atom-coupling parameter around $\lambda\sim10^{5}s^{-1}$in
the microwave domain, this condition can be obtained by adjusting
either the laser and the atomic transition frequencies, $\omega_{l}$
and $\omega_{0}$, or the laser intensity $\Omega$, since the mechanical
frequency $\nu$, being much lesser than the electromagnetic ones,
is irrelevant here. In a similar way, to engineer the Hamiltonian
Eq.(26), a bi-dimensional ion trap is required, and the particle and
anti-particle simulation is encoded in the vibrational states of the
ion in the $x$ and $y$ directions. The condition $\frac{\eta^{2}}{\delta}\tau=\pm\frac{\pi}{2}$
can easily be satisfied by adjusting the external laser frequencies
$\omega_{x}=\omega_{y}=\omega$, the ion internal transition frequency
$\omega_{0}$, the ion vibrational frequencies $\nu_{x}=\nu_{y}=\nu$
in order to obtain a small detuning $\pm\delta=\omega-\omega_{0}-\nu$,
and the laser pulse duration $\tau$. On the other hand, to engineer
the Hamiltonian Eq.(30) that simulates the fermion charge-conjugation
operator, Eq.(31), the internal states of two identical two-level
ions, trapped into the same cavity, are needed. The condition to be
matched is that now the strength of the coupling $\lambda$, the pulse
duration $\tau$ and the detuning between the cavity mode frequency
$\omega$ and the internal transition frequency $\omega_{0}$ of the
ions obey $\frac{\lambda^{2}}{\delta}\tau=\pm\pi/2$, which, although
feasible, is indeed a more stringent condition.

\section{Conclusions}

In this paper we have proposed a method to engineer the unitary charge
conjugation operator as known from quantum field theory for bosons
as well as for fermions \cite{Greiner book}. Although easily extendable
to other contexts such as cavity QED \cite{Fabiano}, we focus on
the highly controllable scenario of trapped ions where quantum controls
of single ion states are daily being reported \cite{Zipkes10}, thus
opening the possibility of simulating particle and antiparticle charge
conjugation. To engineer the bosonic charge-conjugation operator,
we relies on two method: the first one uses both a single mode of
a vibrational ion state and the single mode of a cavity field state,
where the ion is trapped; the second method uses the vibrational harmonic
states of a single trapped ion in two different directions. To engineer
the charge-conjugation operator for fermions, we propose a scheme
which is based on two two-level ions trapped into the same single-mode
cavity, such that the fermionic operators are simulated by the pseudo-spin
operator related to the internal states of the ions.

\section{Acknowledgment}

The author acknowledge financial support from the Brazilian agency
CNPq and Dr. Juan Mateos Guilarte for the kind hospitality during
the stay in USAL. This work was performed as part of the Brazilian
National Institute of Science and Technology (INCT) for Quantum Information.

\end{document}